# Controlling suction by vapour equilibrium technique at different temperatures, application to the determination of the water retention properties of MX80 clay

**Anh-Minh Tang and Yu-Jun Cui**

**CERMES-ENPC, Institut Navier, France**

**Corresponding author**

Dr. Yu-Jun Cui
ENPC/CERMES
6 et 8, av. Blaise Pascal
Cité Descartes, Champs-sur-Marne
77455 Marne La Vallée cedex 2
France
Tel: 33 1 64 15 35 50
Fax : 33 1 64 15 35 62
E-mail : cui@cermes.enpc.fr

# Controlling suction by vapour equilibrium technique at different temperatures, application to the determination of the water retention properties of MX80 clay

**Anh-Minh Tang and Yu-Jun Cui**

**Abstract**

Problems related to unsaturated soils are frequently encountered in geotechnical or environmental engineering works. In most cases, for the purpose of simplicity, the problems are studied by considering the suction effects on volume change or shear strength under isothermal conditions. Under isothermal condition, very often, a temperature independent water retention curve is considered in the analysis, which is obviously a simplification. When the temperature changes are too significant to be neglected, it is necessary to account for the thermal effects. In this paper, a method for controlling suction using the vapour equilibrium technique at different temperatures is presented. First, calibration of various saturated saline solutions was carried out from temperature of 20°C to 60°C. A mirror psychrometer was used for the measurement of relative humidity generated by saturated saline solutions at different temperatures. The results obtained are in good agreement with the data from the literature. This information was then used to determine the water retention properties of MX80 clay, which showed that the retention curve is shifting down with increasing of temperature.

*Key words*: vapour equilibrium technique, hygrometer, water retention curve, temperature effects, compacted bentonite.

**Introduction**

In most geotechnical and environmental works, especially in nuclear waste disposal, temperature effects on the thermo-hydro-mechnical behaviour of unsaturated soils are significant. In analysis of fluid flow, however, retention curves are often taken as being temperature independent, which results in errors. Thus, the present work aims at developing an experimental method for the determination of the water retention curves of heavily compacted swelling clays with controlled suction at different temperatures.

There are three well known suction controlled techniques: i) axis translation technique for suctions up to 14MPa, which is based on the use of ceramic porous stones allowing application of air and water pressure separately (Richards 1941), ii) osmotic technique for suctions up to 12 MPa (Delage et al. 1998), which is based on the use of PEG (polyethylene glycol) solutions, iii) vapour equilibrium technique for suctions ranging from 3MPa to 1000MPa (Tessier 1984; Romero 1999; Delage and Cui 2000; Villar 2000), which is based on the use of salt solutions. Because of the high activity of swelling clays, the vapour equilibrium technique was used in the present work.

A common practice of the vapour equilibrium technique is to place soil samples in a desiccator containing a salt solution, and the samples being supported by a rigid grid above the salt solution (Kanno and Wakamatsu 1993; Delage et al. 1998; Saiyouri et al. 2000; Graham et al. 2001; Romero et al. 2001; Leong and Rahardjo 2002). It is generally observed that the time taken by soil samples to reach equilibrium is about one to two months. Marcial et al. (2002) developed a method to shorten the equilibrium time. They determined the water retention properties of three swelling clays (MX80, Kunigel-V1 and FoCa7) from slurry state by following drying paths. Two procedures were applied. In the first procedure, soil samples were put in a desiccator totally closed, which contained a given saturated saline solution at the bottom. In the second procedure, a pneumatic pump was used to ensure the circulation of humid air generated in a separated bottle by using saturated saline solution at its bottom. This humid air was circulated through the cell that contained the soil samples. The comparison of these two procedures has clearly shown that humid air circulation reduced the equilibrium time significantly, from several months down to 2-4 weeks.

Application of the vapour equilibrium technique in mechanical tests was done by several authors. Belanteur et al. (1997), Villar (1999), Cuisinier and Masrouri (2002) carried out oedometric tests with controlled suction by putting samples in a cell containing salt solutions. As mentioned above, equilibrium time in this case was quite long (i.e., more than three weeks). Bernier et al. (1997), Cui et al. (2002), Lloret et al. (2003), Macial (2003) improved the water moisture exchange between humid air and soil sample in the oedometer by allowing circulation of humid air through the base of sample. This improvement shortened the equilibrium time to a week (Bernier et al. 1997). The triaxial cell designed by Lingnau (Lingnau et al. 1996) was modified to control suctions (Blatz and Graham 2000). The authors allowed humid air circulation around the sample, thus decreasing the drainage length and accelerating water moisture exchanges. In the case of a sand/bentonite mixture, with samples of 50 mm in diameter and 100 mm high, the equilibrium time was about 12 days.

Saturated salt solutions are generally used in the vapour equilibrium technique (Delage et al. 1998; Saiyouri et al. 2000; Loiseau 2001; Montes-H et al. 2003). The advantage of using saturated solution is that the molar fraction of water in a solution does not change with humidity exchanges between liquid phase and gaseous phase. The imposed suction which is related to the molar fraction of water is therefore maintained constant. However limitation of this method to practical application is the discontinuity of the suction values which can be obtained. Usually solutions at different concentrations of sodium chloride (NaCl) or potassium chloride (KCl) are used for small suction range (Fredlund and Rahardjo 1993; Leong and Rahardjo 2002). Others solutions such as sulphuric acid can be also used (Villar 2000; Romero 1999). The use of unsaturated solutions has the advantage to obtain continuous range of suction values, but the difficulty is to keep their concentration constant during the duration of the test. In the present work, only saturated solutions were used.

The vapour equilibrium method was applied at high temperature by Romero (1999) and Villar (2000). They used sulphuric acid solutions at various concentrations to control the suction. The relative humidity was calculated as a function of temperature and the molality of solute by an empirical expression. For other common salt solutions, the majority of data available was given by Schneider (1960). Multon et al. (1991) noted some significant disagreement between different sources from the literature. Therefore a check should be made before any use.

The work reported in the present paper examined a method that allows both suction and temperature to be controlled. Several salt solutions were calibrated in terms of suction and temperature. The suction was measured by using a mirror psychrometer and the temperature was maintained constant by immersing containers in constant temperature water bath. This information was then used to determine the water retention properties of compacted MX80 clay.

## 2. Calibration of salt solutions

2.1. Method

Five pure salts were selected for the present study. The chemical components of these salts are presented in Table 1. As can be seen for most of the salts, the assay value was higher than 99%. Here, assay value is an indication of the purity of the salt used. In practice, there is no completely pure salt. Sodium chloride, for instance, contains not only ions of Na and Cl but also ions of Br, I, Ca, K, N, P, Pb and S. This impurity obviously renders the chemical potential or suction calculations based on the molar fraction of solute impossible. It was therefore necessary to calibrate the salt before its use.

A hygrometer, Hygro-M2, was used to measure the relative humidity (RH) generated by different saturated solutions of salt at various temperatures. Figure 1 shows a schematic view of the experimental system. The hygrometer detects the presence of condensed water on the mirror by a reduction of mirror's reflection. The photodetectors are arranged in an electrical bridge circuit which can control the direct current to the thermoelectric cooler. Firstly, when the mirror is dry, the thermoelectric heat pump cools the mirror toward the dew point. As dew begins to form on the mirror, the mirror's reflectivity decreases which causes a decrease in the cooling current. The system stabilises when a thin dew layer is maintained on the mirror surface. An accurate thermometer embedded within the mirror directly monitors the temperature of the mirror at this condition. This temperature is called the "dew point temperature". More details are described in Loiseau (2001). The specifications of the dew point sensor selected are presented in Table 2.

The experimental set-up for the determination of RH generated by salt solutions is shown in Figure 2. The dew point sensor and the air temperature sensor were introduced in an enclosed metallic cell that was totally immersed in water bath. At the bottom of this cell, a glass recipient containing saturated salt solution was placed. The bath was covered by a layer of polystyrene to reduce the heat exchange between water and environment. A thermostat pump was used to maintain temperature of water in the bath. The variation of water temperature was about ±0.02°C. During a test, dew point temperature $T_{dp}$ related to the saturated salt solution used was measured by the dew point sensor, and cell temperature $T_a$ was monitored by the air temperature sensor. The relative humidity (RH) generated by a salt solution was determined using the following relationship (Wasmer 1988):

$$[1] \quad RH = \frac{P_{Tdp}}{P_{Ta}} = \frac{A.exp\left(B.\frac{T_{dp}}{T_{dp}+C}\right)}{A.exp\left(B.\frac{T_a}{T_a+C}\right)}$$

where A, B and C are constants at atmospheric pressure, and $P_{Tdp}$ et $P_{Ta}$ are saturation vapour pressure at dew point temperature and air temperature, respectively. At atmospheric pressure, A = 6.1078, B = 17.2694, C = 238.3.

Five saturated salt solutions were tested: $MgCl_2$, $Mg(NO_3)_2$, $NaNO_3$, NaCl and KCl. First, after the set-up installation, the thermostat pump was switched on to maintain the bath at a temperature slightly more than ambient temperature (about 20°C). After the stabilisation of temperature and relative humidity, the pump was adjusted to heat the bath to 7°C higher. After the stabilisation of RH at this new temperature, this procedure of heating was repeated every 7°C until 60°C which is the limit value of the hygrometer used.

## 2.2. Results and discussion

The results obtained for the five salt solutions are presented in Figures 3 – 7, in terms of variations of temperature and relative humidity as a function of time. For $MgCl_2$ solution (Figure 3), it is observed that after each adjustment of the thermostat pump, the temperature increased quickly and became stable after about 30 minutes. The relative humidity decreased with the increase of temperature. For the first stage, RH value reached equilibrium after about two hours whereas for others stages the time was considerably less.

Figure 4 shows the results for $Mg(NO_3)_2$ solution. The temperature in the cell, as for $MgCl_2$ solution, reached equilibrium quickly. However, RH equilibrium took a longer time of about five hours. The RH variation at each temperature increment was about 2%, larger than the variation for $MgCl_2$ solution. The initial quick RH change can also be observed in Figure 5 for $NaNO_3$ solution. Compared to $Mg(NO_3)_2$ solution, the RH equilibrium time was shorter because less than three hours was necessary for a variation of 2%.

The results of NaCl solution are presented in Figure 6. It appears that the temperature effect was quite small. After the first stabilisation, each temperature increase gave rise to a very slight increase of RH. This increase in RH is difficult to understand, and it could be related to the chemical property of NaCl salt in term of solubility changes with temperature. Figure 7 presents the result of KCl solution. The RH equilibrium after each temperature change was reached as quickly as $NaNO_3$ solution, i.e. less than three hours for a variation of 2%.

It is also observed that at the beginning of each temperature change, RH value abruptly changed then came back to normal values. This was the case for all the tests. This phenomenon is related to the automatic adjustment of hygrometer.

Figure 8 summarises all results in terms of RH changes with temperature increase, together with the results of Schneider (1960) and Multon et al. (1991). The results of $(NH_4)_2SO_4$, $K_2SO_4$ and $K_2CO_3$ solutions reported in theses references are also plotted because these solutions will be used for the study on water retention properties of MX80 clay. As a whole, results obtained from the present study are very consistent with the Schneider's and Multon et al.'s ones. It appears that RH decreased when temperature was increasing, and that the RH decrease was not at a same rate for all salts. $Mg(NO_3)_2$ has a rate as large as 0.45% / °C. Whereas NaCl has a rate close to zero (after the data of Schneider 1960 and Multon et al.1991). The data of the present study even shows a very slight increase of RH generated by NaCl solution with increasing temperature.

The good agreement of the results obtained with that in the literature shows that the experimental set-up used, with a thermostat pump and polystyrene isolation, is reliable for ensuring the isothermal condition in the RH range considered. The results of Schneider (1960) at 80°C temperature or with other salts can be then used directly.

## 3. Water retention properties of MX80

### 3.1. Material and method

Industrial bentonite MX80 clay was used in the present study. This material contains about 80% montmorillonite. Its properties collected from literature are presented in Table 3. This is a high plastic clay bentonite that is usually studied as a possible engineering barrier for high level nuclear waste disposal. Prior to any tests, the clay was put in a hermetic chamber where relative humidity was controlled at 44% (this corresponds to a suction value of 110 MPa at 20°C) by allowing vapour circulation of saturated $K_2CO_3$ solution. Equilibrium was reached after about 2 months. This equilibration process decreased the clay water content from 17% (at laboratory condition) to 8.5%. The clay was then compacted in a mould to a maximal pressure of 39 MPa to obtain a dry unit weight of 1.7 kN/m$^3$. Sixty samples were compacted in this fashion. The dimensions of the mould were 8.0 mm high and 20.0 mm in diameter. After compaction, samples were taken out of mould and placed in the RH controlled hermetic chamber for more than one week for suction homogenisation. The final dimensions were about 8.7 mm high and 20.2 mm in diameter, which correspond to a dry unit weight of about 16.5 kN/m$^3$.

To study the water retention properties at 20°C, four desiccators which contained three samples in each were placed in an air-conditioned room with temperature controlled at 20±0.5°C. Each desiccator was connected with a bottle containing a saturated salt solution. A peristaltic pump was used to allow the circulation of humid air from the bottle to the desiccator. This experimental set-up is similar to that used by Marcial (2003).

The water retention properties at 40°C, 60°C and 80°C were studied without using circulated air. The compacted soil samples were first introduced in a sealed desiccator containing a saturated salt solution at its bottom. Each desiccator accommodated 3 samples (see Figure 9). Five desiccators were placed in water bath that had been used for salt solutions calibration. The temperature of the bath was maintained at 40±0.1°C during the entire test by using a thermostat. The other desiccators were placed in ovens at temperatures controlled at 60±0.1°C and 80±0.1°C.

The water content of samples was monitored by weighing then every 3 days until the weight stabilised. This was performed in less than 15 s after taking a sample out of desiccator. This duration was considerably short enough to neglect water evaporation during weighing process. The water content at equilibrium was finally determined by oven-drying the sample at 150°C for 24h. Tessier (1984) showed that this drying procedure with 150°C temperature was necessary to determine the water content of swelling clays.

The method, without using circulated air adopted for 40°C, 60°C and 80°C temperature, is based on the consideration of the high temperature gradient between the cell and the room, which can give rise to the condensation phenomenon, and therefore affecting the controlled suction. Bernier et al. (1997) allowed air circulation to impose suction in soil samples at high temperature up to 60°C by introducing the entire circulation system including soil samples holders, salt solutions, pump, relative humidity sensor in a temperature controlled chamber. This obviously corresponds to a costly experimental system. Nevertheless, in our knowledge, the normal pump doesn't work at a temperature higher than 60°C. This would be the reason why Bernier et al. (1997) did not carry out the test at higher temperature.

Because a temperature change will result in a suction change, the accuracy problem related to control of temperature obviously affects the quality of control of suction. Fredlund and Rahardjo (1993) noted that a controlled temperature environment of ±0.001°C was required in order to measure total suctions with an accuracy of ±10 kPa. The information obtained from the salt solutions calibration indicated that a temperature change of ±0.5°C (the maximum accuracy of temperature controlled for isothermal water retention tests) would not involve a significant change in RH. However, an abrupt change of temperature may always cause an abrupt change of RH. In the case of high RH environment, this abrupt change may involve the water condensation.

The method without air circulation certainly minimises the risk of condensation because the temperature gradient is much less in comparison with the case of air circulated system. Nevertheless, this method increases the equilibrium time. It is evident that this problem can be compensated by increasing the contact surface between soil samples and humid air or by reducing sample mass. In this regard, Montes-H (2002) had an equilibrium time less than one week with a MX80 sample of 1 g.

### 3.2. Results and d0iscussion

In the subsequent analysis, the term "suction" will be used. The conversion from RH to suction is obtained by using equation [2] (Fredlund and Rahardjo 1993):

[2] $$s = -\frac{\rho_W RT}{M_W} \ln\left(\frac{RH\%}{100}\right)$$

Where:

$s$ is the soil suction (kPa)

$R$ is the universal (molar) gas constant ($R = 8.31432$ J/mol.K)

$T$ is the absolute temperature ($T = 273.16 + t$)

$t$ is the temperature (°C)

$M_W$ is the molecular mass of water vapour ($M_W = 18.016$ g/mol)

$\rho_W$ is the volumetric mass of water (kg/m$^3$).

Accordant to the data of Incropera and De Witt (1996), for T = [273.16°K, 365°K]:

$\rho_W = 1.10^{-10}.T^6 - 2.10^{-7}.T^5 + 2.10^{-4}.T^4 - 0.085.T^3 + 20.26.T^2 - 2565.T + 135464$ (kg/m$^3$).

Figure 10 shows the variation of water contents with time at 20°C. As mentioned above, three samples were tested at each suction value. For 145 MPa suction, all three samples were performed in a similar fashion. A slight decrease of water content indicates that the initial suction value of the samples was lower than 145 MPa. For 82 MPa suction, a quick increase in water content was observed which was then followed by a decrease towards an equilibrium stage. This response was pronounced particularly at suction value of 20 MPa. This phenomenon can be explained by a possible leakage in the air circulation. During the tests, the results were frequently compared with the available data of Marcial (2003). It was concluded that the water contents at peak stage are unacceptably high. The experimental set-up was then checked carefully to detect any possible leakage. After overcoming this particular problem, the water exchange was brought to the normal form. For 37 MPa suction, it is observed that stabilisation was quickly attained (i.e., 10 days), and that, however, an abrupt increase took place at about 22 days. This phenomenon is again related to a technical problem: the plastic tube in peristaltic pump was twisted, interrupting humid air circulation. Thus, the observed abrupt increase of water content corresponds to the resolution of the problem. The final values of water content were related to the relevant values of suction. The results obtained are in good agreement with that of Marcial (2003).

Figure 11 shows the water content variations at 80°C, in the case of oven-heating, with 5 suction values: 8 MPa (K$_2$SO$_4$), 39 MPa (KCl), 47 MPa (NaCl), 67 MPa (NaNO$_3$) and 184 MPa (Mg(NO$_3$)$_2$). A peak value is also observed on two samples of 8 MPa suction, which was due to the dropping down of condensed water from the desiccator's cover. There was not any condensed water observed on the third sample in this desiccator because the two recipients containing the two other samples covered it (see Figure 9 for the experimental setup). For the other samples in the other desiccators at higher suction values, as the RH was far from the saturation (100%), this problem of water condensation was not observed. For 67 MPa suction, no water exchange was observed, showing that the initial suction of the soil at 80°C was close to this value.

The results are gathered in Figure 12, together with that obtained by using other salt solutions and at other temperatures. In this figure, each point corresponds to a suction value, a temperature and the average value of the final water content from the three identical tests. The two regression curves was added to the isothermal water retention curves at 20°C and 80°C. It was observed that an increase of temperature gives rise to an increase of suction in the desiccator, thus a decrease of water content in the sample with the same salt solution. Nevertheless, the shifting down of the water retention curve at 80°C from that at 20°C shows that the temperature has influenced the water retention capacity of clay.

The entire data was plotted in the Figure 13. Four regression curves was added to clarify the temperature effects on the water retention curve of this clay. It appears that when temperature was increasing the curve shifted down. That means at the same suction value, the water content of the sample at higher temperature was less than that at lower temperature. In another way, the increase of temperature decreased the water retention capacity of the sample. For a given water content, an estimation of suction change rate with increasing temperature ($\Delta \lg s/\Delta T°$) was approximately $-2.9.10^{-3}$ (lgMPa/°C). Other researchers carried out similar studies on temperature effects on water retention curves of compacted clay. Some of these studies are cited here: Romero et al. (2001) and Bernier et al. (1997) on Boom clay; Olchitzky (2002) on French FoCa7 bentonite; Kanno and Wakamatsu (1993) on Kunigel-V1 bentonite;

Villar (2000) and Pintado (2002) on "Cabo de Gata" bentonite. They all found that the temperature effects on the water retention curve were very slight.

Olchitzky (2002) tried to explain the temperature effect by the interfacial tension decrease with temperature increase and concluded that the interfacial tension decrease was not sufficient to explain the experimental results. In the present study, similar analysis was made by using the data reported by Fredlund and Rahardjo (1993). According to Jurin's law (equation [3]), capillary suction is a function of capillary radius:

$$[3] \quad u_a - u_w = \frac{2T_s \cdot \cos\theta}{r}$$

where:

$u_a$-$u_w$ is the capillary suction (Pa).

$T_s$ is the interfacial tension between water and air (N/m).

$T_s$ (mN/m) = -2,707.$10^{-4}$ .$t°^2$ –0,1420.$t°$ + 75,69

$t°$ is the temperature (°C).

$\theta$ is the contact angle, generally assumed to be zero.

$r$ is the radius of the capillary tube (m).

Figure 14 plots the capillary suction against pore size radius for two values of temperature (20°C and 80°C). For a given radius of capillary tube, a decrease of suction was obtained with the temperature increase. If we simply suppose that the temperature change had no effect on the geometric configuration of the soil water system, with the same water content, the pore radius of the air-water surface in soil at difference temperatures should be the same. In that case, the difference in suction defined by the distance between the two straight lines for a given pore radius can be compared with the obtained experimental values. This difference is estimated to be -1.1 $10^{-3}$ (lgMPa/°C), which is comparable with the experimental value mentioned above (-2.9.10$^{-3}$ lgMPa/°C). Because the Jurin's law only involves the capillary suction, the good consistence between theoretical estimation and experimental result shows that in the range of suction considered, a temperature elevation decreased the interfacial tension $T_s$, therefore decreases the capillary suction.

## 4. Conclusion

An experimental set-up was developed in order to calibrate the relative humidity generated by salt solutions at different temperatures. The results obtained from this calibration are in good agreement with that published in literature. This information was then used to determine the water retention properties of MX 80 clay at four different temperatures: 20°C, 40°C, 60°C and 80°C. It was observed that for a given water content, the suction decreased with increasing temperature. That means the temperature elevation has reduced the water retention capacity of soil. The experimental suction decrease rate was compared with that from a theoretical estimation by using Jurin's law and by considering an interfacial tension which changes with temperature. A good consistency was observed, showing that it is the capillary suction which was affected by the temperature changes.

## References

Belanteur, N., Tacherifet, S., and Pakzad, M. 1997. Etude des comportements mécanique, themo-mécanique et hydro-mécanique des argiles gonflantes et non gonflantes fortement compactées. Revue Française de Géotechnique, **78**: 31-50.

Bernier, F., Volkaert, G., Alonso, E., and Villar, M. 1997. Suction-controlled experiments on Boom clay. Engineering Geology, **47**: 325-338.

Blatz, J., and Graham, J. 2000. A system for controlled suction in triaxial tests. Géotechnique **50**(4): 465-469.

Cerato, A.B., and Lutenegger A.J. 2002. Determination of Surface Area of Fine- Grained Soils by the Ethylene Glycol Monoethyl Ether (EGME) Method. Journal of Geotechnical Testing, **25**(3): 314-320.

Cui, Y. J., Yahia-Aissa, M., and Delage, P. 2002. A model for the volume change behaviour of heavily compacted welling clays. Engineering Geology, **64**: 233-250.

Cuisinier, O., and Masrouri, F. 2002. Study of the hydro-mechanical behaviour of a swelling soil under high suctions. *In* Proceeding of the 3$^{rd}$ International Conference on Unsaturated Soils/UNSAT 2002/Recife/BRAZIL, *Edited by* Jucá, de Campos & Marinho, Vol. 2, pp. 587-591.

Delage, P., and Cui, Y.J. 2000. L'eau dans les sols non saturés. Techniques de l'Ingénieur, C301, traité Construction, volume C2.

Delage, P., Howat, M.D., and Cui, Y.J. 1998. The relationship between suction and swelling properties in a heavily compacted unsaturated clay. Engineering Geology, **50**: 31-48.

Fredlund, D.G., and Rahardjo, H. 1993. Soil mechanics for unsaturated soils. John Wiley & Sons, Inc, 517p.

Graham, J., Blatz, J.A., Alfaro, M.C., and Sivakumar, V. 2001. Behavioral influence of specimen preparation methods for unsaturated plastic compacted clays. *In* Proceedings of the 15$^{th}$ International Conference on Soil Mechanics and Geotechnical Engineering, Istanbul, Vol. 1, pp. 633-638.

Incropera, F.P., and De Witt, D.P. 1996. Fundamentals of heat and mass transfer. 4$^{th}$ ed., John Wiley & Sons, Inc. 886 p.

Kanno, T., and Wakamatsu, H. 1993. Moisture adsorption and volume change of partially saturated bentonite buffer materials. Materials Research Society Symposium Proceeding, Vol. 294, pp. 425-430.

Leong, E.C., and Rahardjo, H. 2002. Soil-water characteristic curves of compacted residual soils. *In* Proceeding of the 3$^{rd}$ International Conference on Unsaturated Soils/UNSAT 2002/Recife/BRAZIL, *Edited by* Jucá, de Campos & Marinho, Vol. 1, pp. 271-276.

Lingnau, B.E., Graham, J., Yarechewski, D., Tanaka, N., and Gray, M.N. 1996. Effects of temperature on strength and compressibility of sand-bentonite buffer. Engineering Geology, **41**: 103-115.

Lloret, A., Villar, M.V., Sanchez, M., Gens, A., Pintado, X., and Alonso, E.E. 2003. Mechanical behaviour of heavily compacted bentonite under high suction changes. Géotechnique, **53**(1): 27-40.

Loiseau, C. 2001. Transferts d'eau et couplages hydromécaniques dans les barrières ouvragées. Ph.D. thesis, Ecole Nationale des Ponts et Chaussées, Paris, France.

Marcial, D. 2003. Comportement hydromécanique et microstructural des matériaux de barrière ouvragée. Ph.D. thesis, Ecole Nationale des Ponts et Chaussées, Paris, France.

Marcial, D., Delage, P., and Cui, Y.J. 2002. On the high stress compression of bentonites. Candadian Geotechnical Journal, **39**: 812-820.

Mata, C., Romero, E., and Ledesma, A. 2002. Hydro-chemical effects on water retention in bentonite-sand mixtures. *In* Proceeding of the 3$^{rd}$ International Conference on Unsaturated Soils/UNSAT 2002/Recife/BRAZIL, *Edited by* Jucá, de Campos & Marinho, Vol. 1, pp. 283-288.

Montes-H, G. 2002. Etude expérimentale de la sorption d'eau et du gonflement des argiles par microscopie électronique à balayage environnementale (ESEM) et l'analyse digitale d'images. Ph.D. thesis, Université Louis Pasteur, Strasbourg, France.

Montes-H, G., Duplay, J., Martinez, L., and Mendoza, C. 2003. Swelling-shrinkage kinetics of MX80 bentonite. Applied Clay Science, **22**: 279-293.

Multon, J.L., Bizot, H., and Martin, G. 1991. Chapitre 1: Mesure de l'eau absorbée dans les aliments. *In* Technique d'analyse et de contrôle dans les industries agro-alimentaires. *Edited by* Techniques et Documentation, Lavoisier, Paris, Vol. 4, pp. 1-63.

Olchitzky, E. 2002. Couplage hydromécanique et perméabilité d'une argile gonflante non saturée sous sollicitations hydriques et thermiques. Courbe de sorption et perméabilité à l'eau. Ph.D. thesis, Ecole Nationale des Ponts et Chaussées, Paris, France.

Pintado, X. 2002. Caracterizacion del comportamiento termo-hidro-mecanico de arcillas expansivas. Ph.D. thesis, Universitat Politècnica de Catalunya, Barcelona, Spain.

Richards, L.A. 1941. A pressure-membrane extraction apparatus for soil solution. Soil science, **51**: 377-386.

Romero, E. 1999. Characterization and thermo-hydro-mechnical behaviour of unsaturated Boom clay: An experimental study. Ph.D. thesis, Universitat Politècnica de Catalunya, Barcelona, Spain.

Romero, E., Gens, A., and Lloret, A. 2001. Temperature effects on the hydraulic behaviour of an unsaturated clay. Geotechnical and Geological Engineering, **19**(3-4): 311-332.

Saiyouri, N., Hicher, P.Y., and Tessier, D. 1998. Microstructural analysis of highly compacted clay swelling. *In* Proceedings of the 2$^{nd}$ International Conference on Unsaturated Soils/UNSAT '98/Beijing/CHINA, Vol. 1, pp. 119-124.

Saiyouri, N., Hicher, P.Y., and Tessier, D. 2000. Microstructural approach and transfer water modelling in highly compacted unsaturated swelling clays. Mechanics of cohesive-frictional materials, **5**: 41-60.

Schneider, A. 1960. Neue Diagramme zur Bestimmung der relativen Luftfeuchtigkeit über gesättigten wässerigen Salzlösungen und wässerigen Schwefelsäurelösungen bei verschiedenen Temperaturen. Zeitschrift HOLZ als Roh- und Werkstoff, **18**: 269-272.

Tessier, D. 1984. Etude expérimentale de l'organisation des matériaux argileux: Hydratation, gonflement et structuration au cours de la dessiccation et de la réhumectation. Thèse de doctorat d'état. Université de Paris VII, Paris, France.

Villar, M.V. 1999. Inverstigation of the behaviour of bentonite by means of suction-controlled oedometer tests. Engineering Geology, **54**: 67-73.

Villar, M.V. 2000. Caracterizacion termo-hidro-mecanica de una bentonita de Cabo de Gata. Ph.D. thesis, Universidad Complutense de Madrid, Madrid, Spain.

Wasmer, C. 1988. Pratique de l'hygrométrie : Notions fondamentales et utilisation des diagrammes. *Edited by* Elcowa, France.

## List of tables

## List of figures

**Table 1.** Salts chemical components

| Component | Salts | | | | | | | |
|---|---|---|---|---|---|---|---|---|
| | $(NH_4)_2SO_4$ | $MgCl_2$ | $Mg(NO_3)_2$ | $KCl$ | $K_2CO_3$ | $K_2SO_4$ | $NaCl$ | $NaNO_3$ |
| Assay (>%) | 99 | 99 | 99 | 99.5 | 99.92 | 99 | 100 | 99.5 |
| Ba (<) | - | - | 20ppm | - | - | - | - | - |
| Br and I (<) | - | - | - | - | - | - | 0.005% | - |
| Ca (<) | 0.005% | 0.020% | 100ppm | 0.003% | 0.97ppm | 0.005% | 4.83ppm | 0.002% |
| Cl (<) | 0.002% | - | 10ppm | - | 20ppm | 0.0005% | - | 0.0005% |
| Cu (<) | 0.0005% | - | - | - | 0.20ppm | - | - | - |
| Fe (<) | 0.001% | 0.0005% | 5ppm | 0.0005% | 0.32ppm | 0.0005% | - | 0.0002% |
| K (<) | 0.02% | - | - | - | - | - | 1.43ppm | - |
| Mg (<) | 0.002% | - | - | 0.001% | 0.20ppm | 0.002% | - | - |
| N (<) | - | 0.015% | - | 0.004% | 4.00ppm | - | 4.00ppm | - |
| Na (<) | 0.02% | - | - | 0.02% | 311.00ppm | 0.02% | - | - |
| P (<) | 0.005% | - | - | - | 1.00ppm | - | 0.08ppm | - |
| Pb (<) | 0.002% | 0.0005% | 5ppm | 0.0005% | 0.40ppm | 0.0005% | 0.10ppm | 0.0002% |
| S (<) | - | - | - | - | - | - | 0.34ppm | - |
| Si (<) | 0.01% | - | - | - | - | - | - | - |
| Sr (<) | - | - | 20ppm | - | - | - | - | - |
| Zn (<) | 0.01% | 0.005% | - | - | 0.200ppm | - | - | - |
| $NO_3$ (<) | 0.01% | - | - | - | - | - | - | - |
| $PO_4$ (<) | - | 0.0005% | - | 0.001% | - | - | - | 0.001% |
| $SO_4$ (<) | - | 0.002% | 20ppm | 0.0002% | - | - | - | 0.003% |

ppm : part per million

**Table 2:** Specifications of dew point sensor

| Specifications | Value |
|---|---|
| Relative humidity measurement range | 0% to 100% |
| Dew point temperature measurement range | -40°C to +60°C |
| Accuracy (complete system at 25°C) | |
| - Dew point | ±0.2°C |
| - Air temperature | ±0.2°C |
| - % RH at 40% | ±0.5% nominal |
| - % RH at 95% | ±1.25% nominal |
| Hysteresis | non |

**Table 3.** Properties of Bentonite MX80 clay

| Properties | MX80 |
|---|---|
| Type | Na-Ca type (2) |
| Montmorillonite content (%) | 80 (3) |
| Particle density (Mg/m$^3$) | 2.76 (4) |
| Liquid limit (%) | 519 (5) |
| Plastic limit (%) | 35 (5) |
| Plasticity index | 484 (5) |
| Activity $^{(*)}$ | 8.01 (5) |
| Clay (<2µm) content (%) | 60 (5) |
| Cation exchange capacity (meq/100g) | 82.3 (1) |
| Exchange capacity of Na$^+$ (meq/g) | 79.8 (1) |
| Exchange capacity of Ca$^{2+}$ (meq/g) | 5.28 (1) |
| Specific surface area, S (m$^2$/g) | 522 (1) |

$^{(*)}$ Ratio of plasticity index to percent clay size
(1) Saiyouri et al. 1998.
(2) Marcial et al. 2002.
(3) Montes-H et al. 2003.
(4) Mata et al. 2002.
(5) Cerato and Lutenegger. 2002.

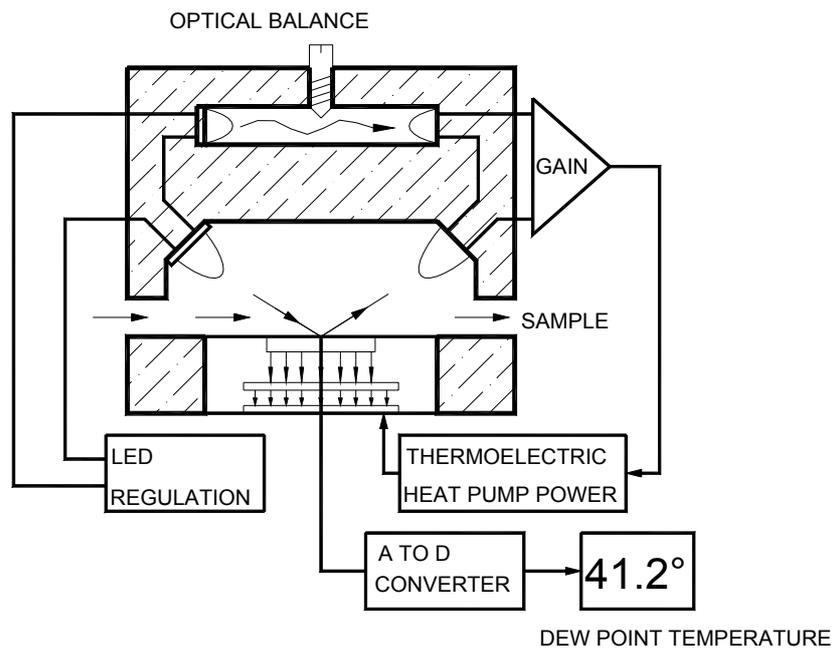

**Figure 1.** Dew point detection in Optical Condensation Hygrometer

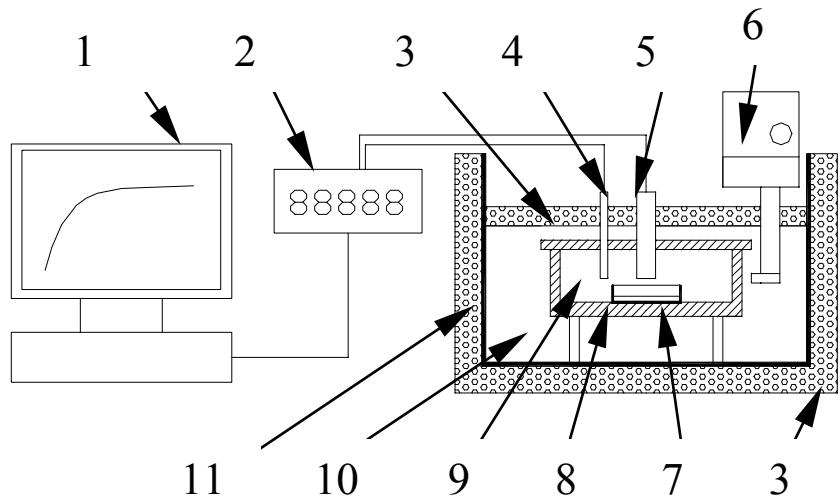

1-Computer
2-Hygrometer M2
3-Polystyrene
4-Temperature sensor
5- Dew point sensor
6-Thermostat pump
7-Salt solution glass cup
8-Metallic cell
9-Humid air
10-Water
11 – Plastic bath

**Figure 2.** Experimental set-up to monitor relative humidity generated by saturated salt solution at different temperatures.

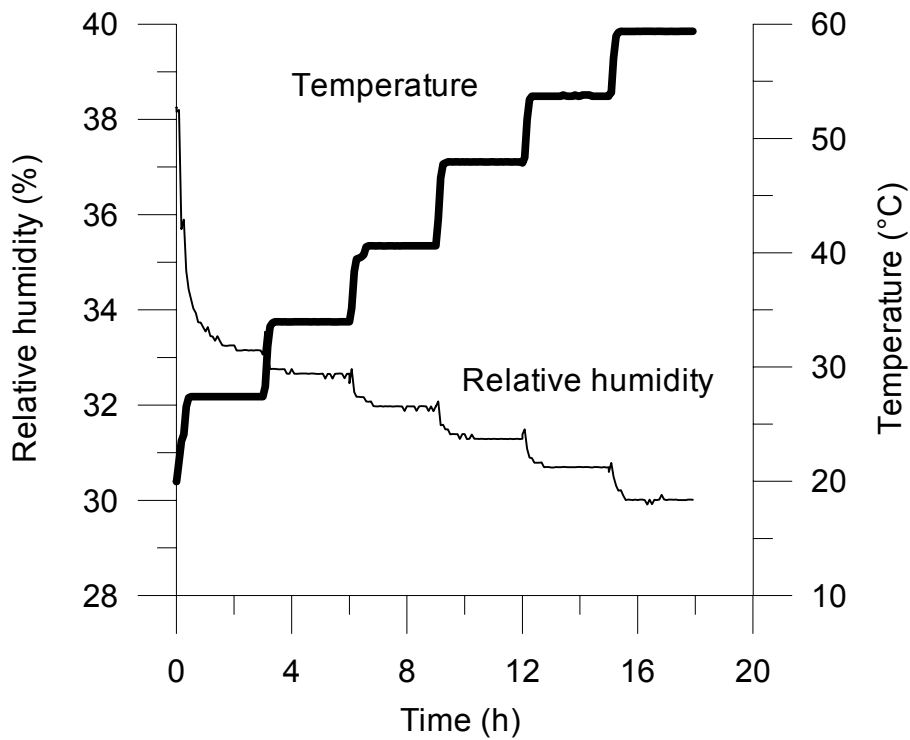

**Figure 3.** Relative humidity measurement of MgCl$_2$ solution

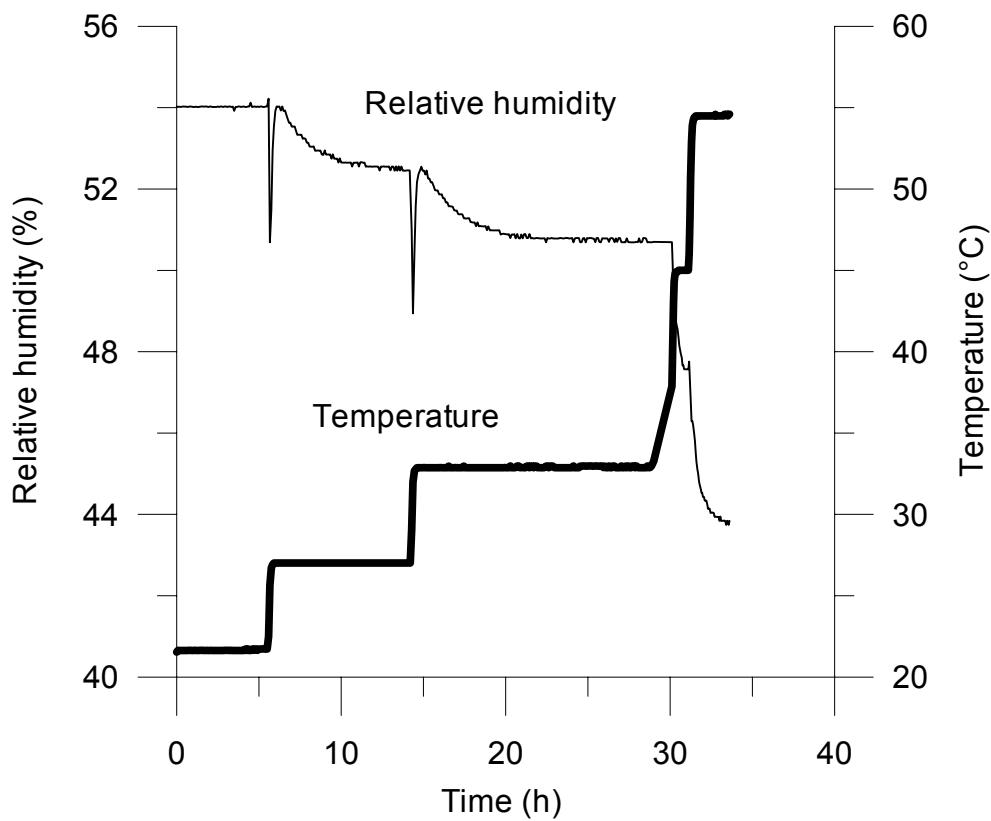

**Figure 4.** Relative humidity measurement of Mg(NO$_3$)$_2$ solution

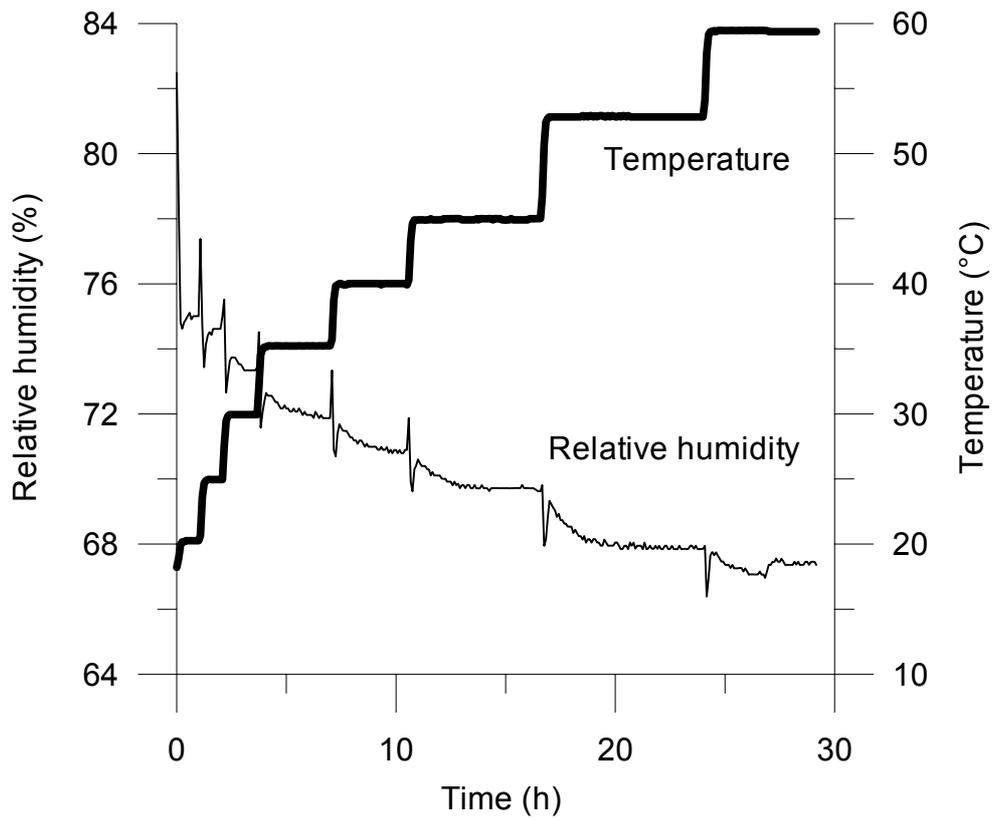

**Figure 5.** Relative humidity measurement of NaNO$_3$ solution

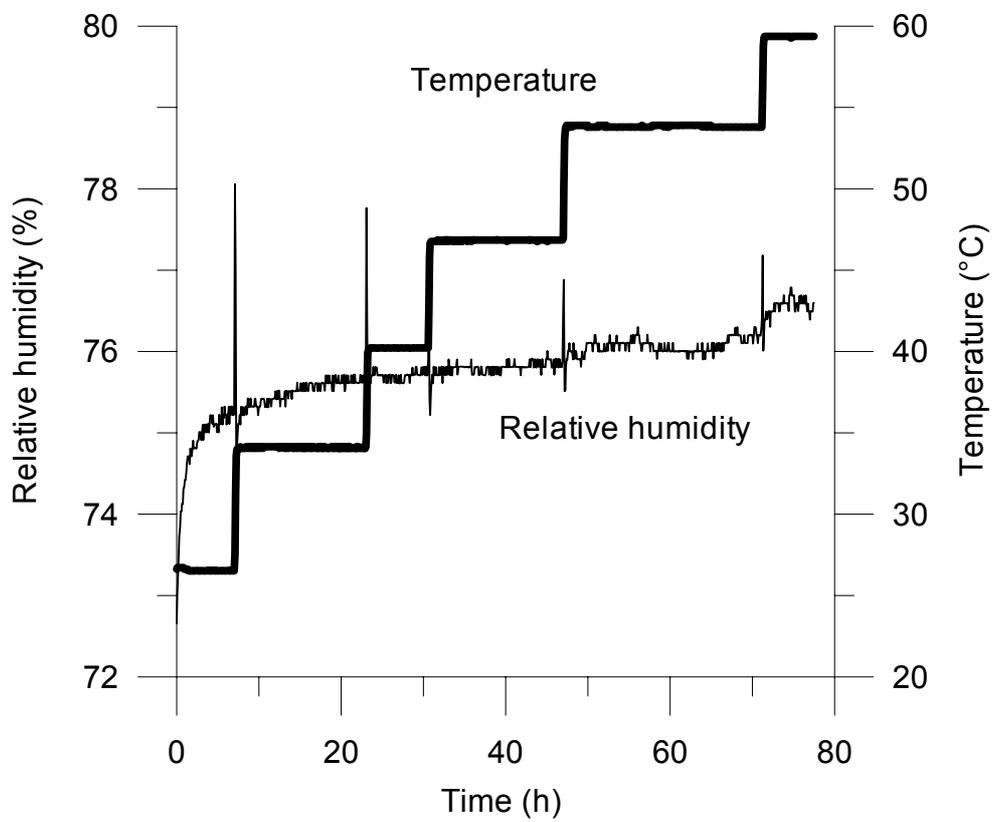

**Figure 6.** Relative humidity measurement of NaCl solution

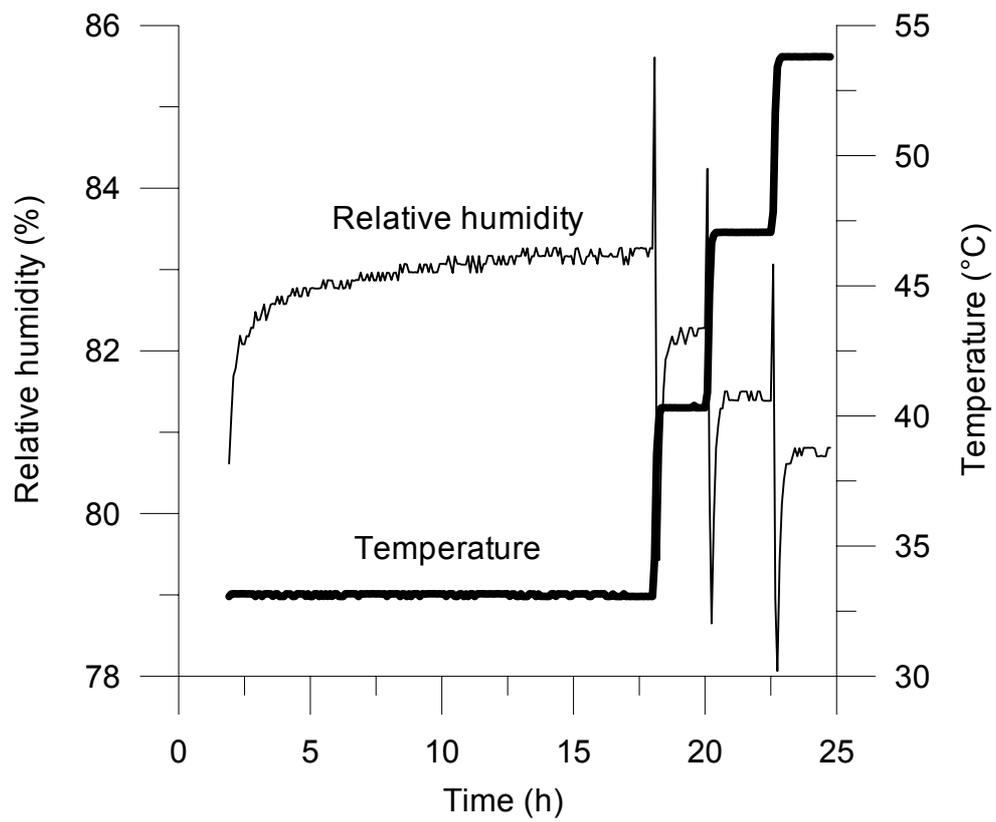

**Figure 7.** Relative humidity measurement of KCl solution

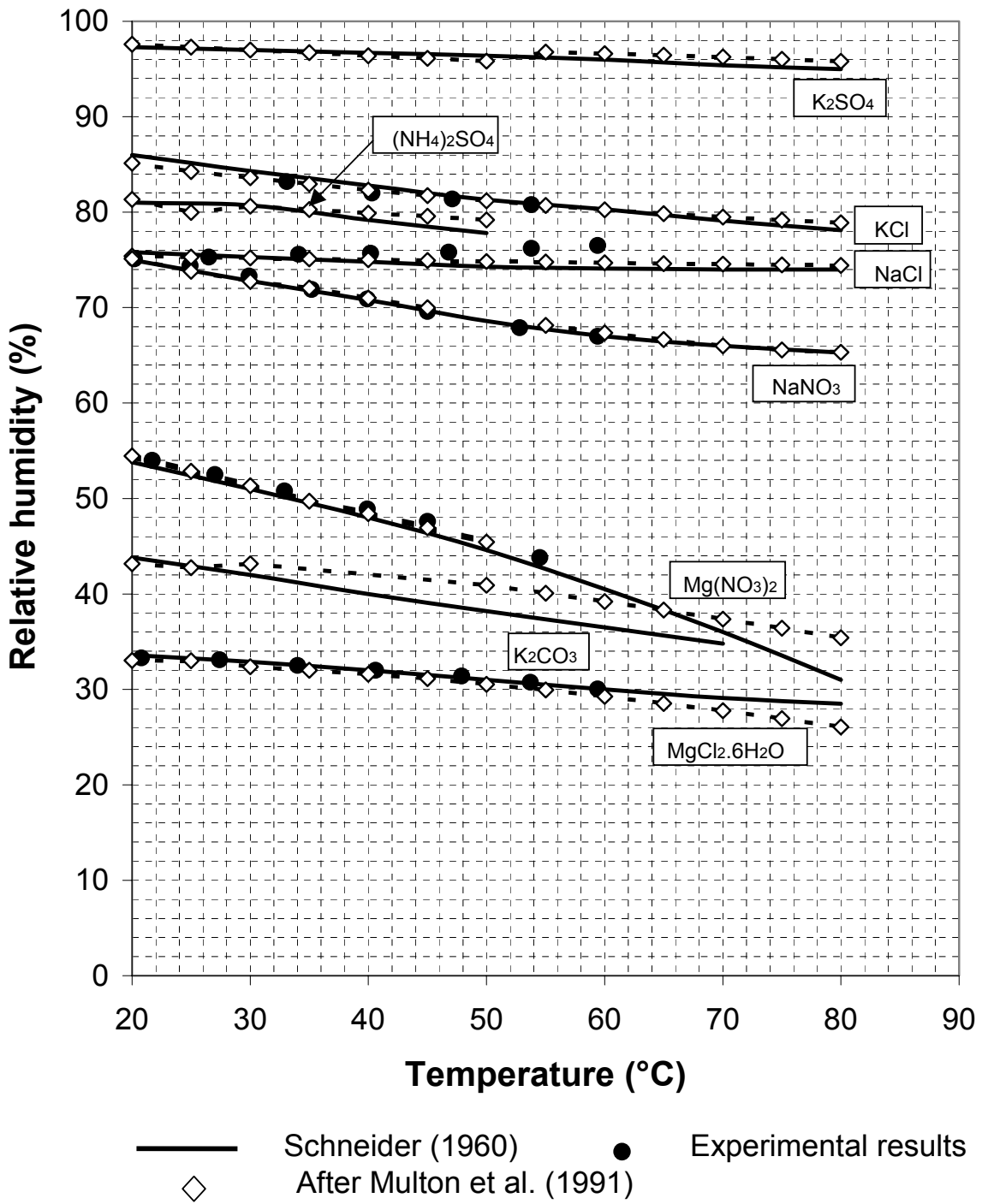

**Figure 8.** Relative humidity change with temperature for different salt solutions

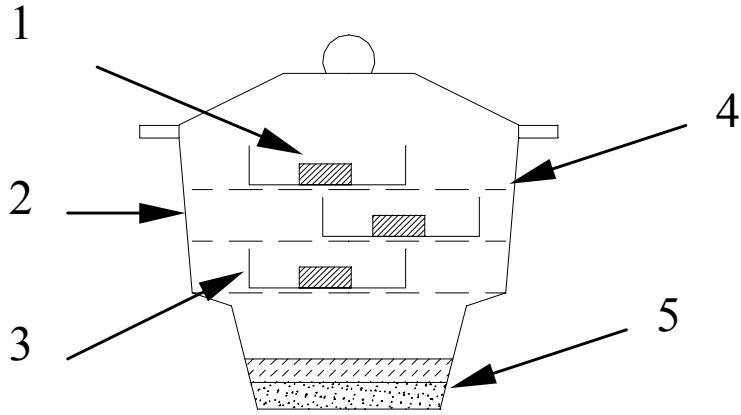

1 – Soil sample
2 – Desiccator
3 – Glass cup
4 – Support
5 – Over saturated salt solution

**Figure 9.** Imposing suction to three samples contained in a desiccator

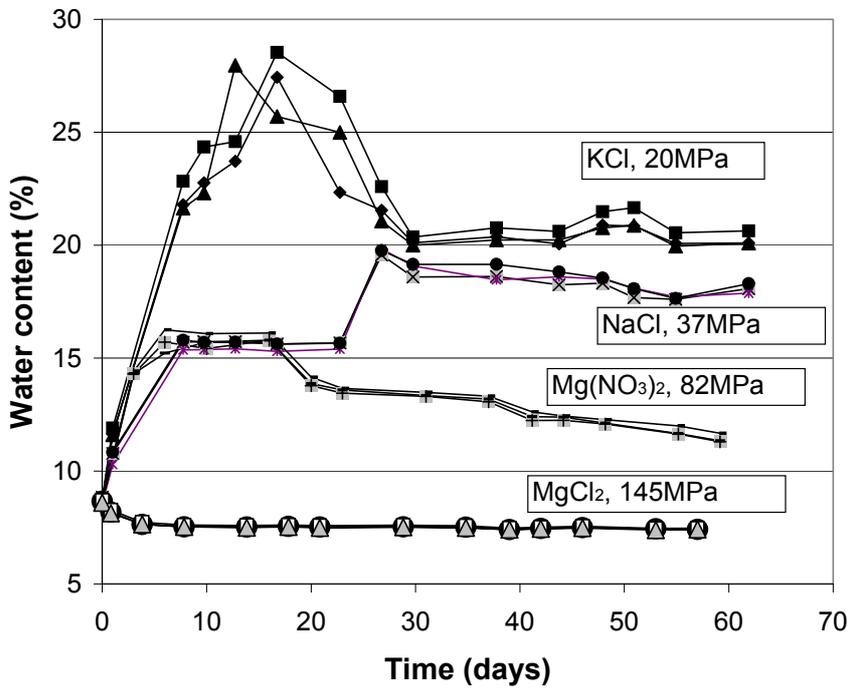

**Figure 10.** Water content variation at 20°C, with air circulation

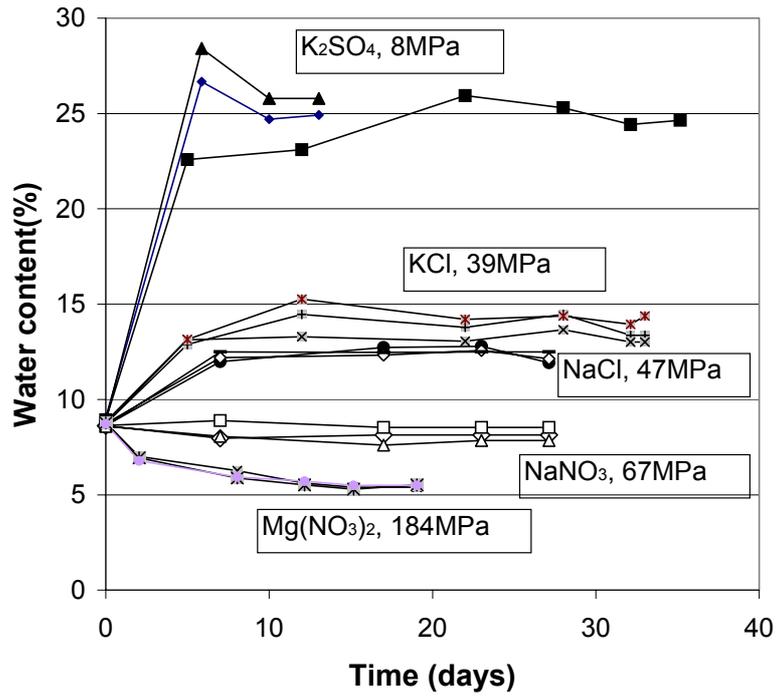

**Figure 11.** Water content variation at 80°C, without air circulation

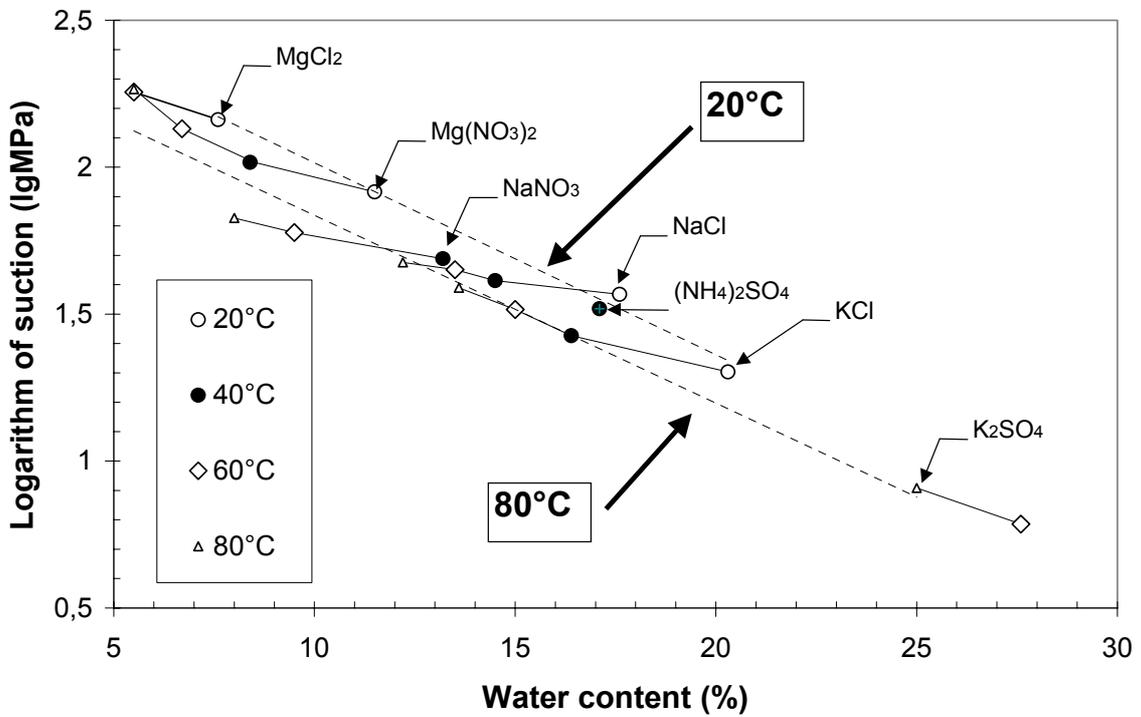

**Figure 12:** Temperature effect on water retention of MX80 clay.

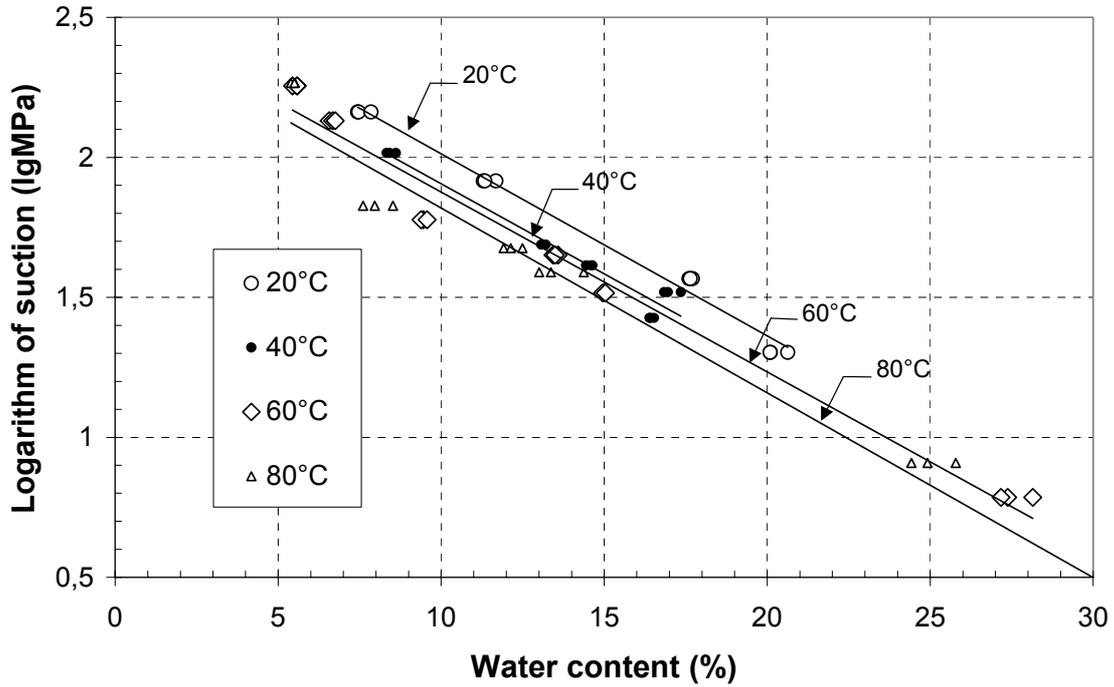

**Figure 13.** Water retention curve at different temperatures

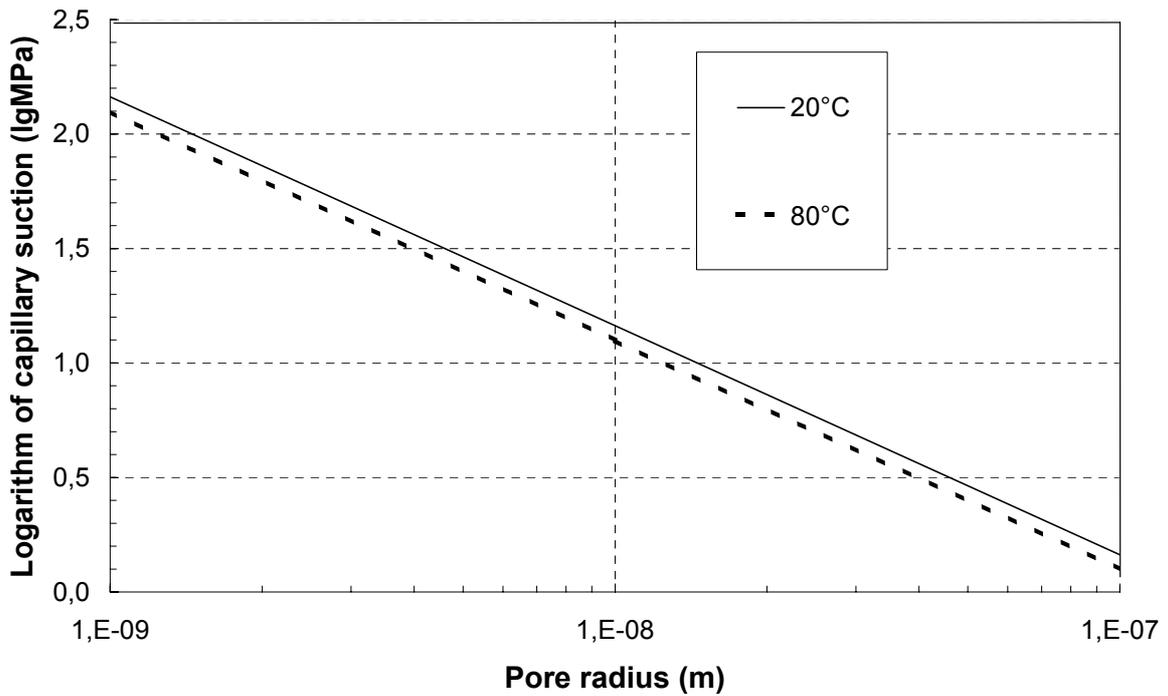

**Figure 14.** Temperature effect on capillary suction through surface tension change